\documentclass[letterpaper,twocolumn,pra,floatfix, showpacs]{revtex4}

\usepackage[utf8]{inputenc}
\usepackage{bm}
\usepackage{multirow,amssymb,amsbsy,amsmath,bbold}
\usepackage{graphicx}
\usepackage{verbatim}
\usepackage{color}
\usepackage{array}
\usepackage{tikz}
\usetikzlibrary{backgrounds,fit,decorations.pathreplacing,decorations.pathmorphing} 
\makeatletter
\usepackage{pifont}
\makeatother

\newcommand{\ket}[1]{| #1 \rangle}
\newcommand{\bra}[1]{\langle #1 |}

\newcolumntype{?}{!{\vrule width 2pt}}
\begin{document}

\title{Non-Markovian dynamics in two-qubit dephasing channels with an application to superdense coding}

\author{Antti Karlsson}
\email{antti.karlsson@utu.fi}
\affiliation{Turku Centre for Quantum Physics, Department of Physics and Astronomy, University of
Turku, FI-20014 Turun yliopisto, Finland}

\author{Henri Lyyra}
\affiliation{Turku Centre for Quantum Physics, Department of Physics and Astronomy, University of
Turku, FI-20014 Turun yliopisto, Finland}

\author{Elsi-Mari Laine}
\affiliation{Turku Centre for Quantum Physics, Department of Physics and Astronomy, University of
Turku, FI-20014 Turun yliopisto, Finland}

\author{Sabrina Maniscalco}
\affiliation{Turku Centre for Quantum Physics, Department of Physics and Astronomy, University of
Turku, FI-20014 Turun yliopisto, Finland}

\author{Jyrki Piilo}
\affiliation{Turku Centre for Quantum Physics, Department of Physics and Astronomy, University of
Turku, FI-20014 Turun yliopisto, Finland}

\date{\today}
\begin{abstract}
We study the performance of two measures of non-Markovianity in detecting memory effects in two-qubit dephasing channels. By combining independent Markovian and non-Markovian noise on the qubits, our results show that the trace distance measure is able to detect the memory effects when at least one of the local channels displays non-Markovianity. 
A measure based on channel capacity, in turn, becomes non-zero when the global two-qubit dynamics shows memory effects.
We apply these schemes to  a well-known superdense coding protocol 
and demonstrate an optimal noise configuration to maximize the information transmission with independent local noises.
\end{abstract}


\maketitle
\section{Introduction}
Quantum information protocols, such as quantum teleportation, quantum cryptography or quantum key distribution \cite{teleportation,BB84}, rely on faithful transmission of quantum information between several parties. However, in any practical scenario, during the transmission, errors take place. The influence of the errors in a quantum information context is described via the theory of noisy quantum channels \cite{Nielsen}. 

Quantum channels describing the noise during transmission of information are often described as "black box'' input-output systems. However, in practice, an interaction with a quantum environment generates the noise \cite{BREUERBOOK} giving rise to a continuous family of channels instead of just one input-output system. Then, the interaction time (or channel length) influences strongly the properties of the channel.  The conventional wisdom is that the noise is harmful for all quantum protocols and isolation from the surroundings is necessary for harnessing the quantum properties. However, recent work has shown, how adding even more noise to the system can actually be beneficial in certain cases \cite{nonlocal,nonlocalmem,expNLNM,entdis,qrelays}. Quantum information protocols, such as entanglement swapping, distillation, quantum teleportation, and quantum key distribution have been shown to benefit from correlated noise, when non-Markovian features are present. 

In recent years there has been rapid progress both in theory  and experimental control of non-Markovian open quantum systems \cite{wolf,NMprl,rivas,mani,determinant,LFS,NMNP,rome1,rome2,brasil}. Further,  first theoretical proposals for exploiting non-Markovianity for quantum information processing and metrology exist \cite{nonlocalmem,metrology,channelcap}. However, many questions related to the proper quantification of non-Markovianity \cite{wolf,NMprl,rivas,mani,determinant,LFS} and to the exploitation of memory effects as a quantum resource still remain elusive. Especially, the additivity  properties of the various non-Markovian measures remain largely unknown \cite{additivity1,additivity2,additivity3}.

The main question in this paper is the following: How, in a practical example, do the different measures describe the dynamics of resources for quantum information tasks? Here we study the case of two independent qubit channels. We first study the non-Markovianity properties of the global channel via two suggested measures for non-Markovian dynamics and then consider an application in superdense coding \cite{bibsdc}; one of the best known examples of using entanglement for quantum information processing purposes. 

\section{Non-Markovian dynamics of two-qubit dephasing channels}

Since the experimental platforms for studying quantum systems allow sophisticated engineering schemes, the importance of non-Markovian processes in open quantum systems has become crucial leading to a vast development towards a general consistent theory of memory effects in quantum dynamics. This has led to an active discussion on the proper definition and quantification of non-Markovian effects in recent years \cite{wolf,NMprl,rivas,mani,determinant,LFS}. Here, we study two information theoretically motivated measures for non-Markovian dynamics and see how well they capture the influence of environment engineering in the performance of superdense coding protocol.

Dynamics of open quantum systems influenced by noise, are described with a family of completely positive, trace preserving (CPTP) maps, denoted $\{\Phi_t\}_{t>0}$. Each member $\Phi_t$ of the family is a CPTP map that evolves an input state $\rho(0)$ from time zero to an output state $\rho(t)=\Phi_t\rho(0) $ at time $t$. 
We study the case where two qubits are subjected to independent and uncorrelated dephasing channels which can be tuned to exhibit both Markovian and non-Markovian dynamics.  This type of dynamics can be experimentally realised with a high degree of environment engineering and further, the model allows to analytically treat most of the non-Markovianity measures.

In matrix form, the evolution of the density matrix of our bipartite system of interest, $\rho^{AB}$, can be written as
\begin{widetext}
\begin{align}\nonumber
\rho^{AB}(t) = 
\begin{pmatrix} 
\rho_{11} & \rho_{12}\kappa_2(t) & \rho_{13}\kappa_1(t) & \rho_{14}\kappa_1(t)\kappa_2(t) \\ 
\rho_{21}\kappa_2^*(t) & \rho_{22} 	& \rho_{23}\kappa_1(t)\kappa_2^*(t) & \rho_{24}\kappa_1(t) \\ 
\rho_{31}\kappa_1^*(t) & \rho_{32}\kappa_1^*(t)\kappa_2(t) & \rho_{33} & \rho_{34}\kappa_2(t) \\ 
\rho_{41}\kappa_1^*(t)\kappa_2^*(t) & \rho_{42}\kappa_1^*(t) & \rho_{43}\kappa_2^*(t) & \rho_{44} 
\end{pmatrix},
\end{align}
\end{widetext}
where $\kappa_i(t)$ are the complex valued decoherence functions having absolute values between 0 and 1. Their time dependence dictates the properties of the quantum channel completely. The terms containing products of the $\kappa_i$ coefficients reflect the fact that the local channels are independent. For general, possibly correlated, channels these products should be replaced with more general functions, containing information about the correlations \cite{nonlocal}. Equivalently, the independence of the local channels means that the dynamical map of the total system is a tensor product of local dynamical maps: $\Phi_t^{AB} = \Phi_t^{A}\otimes \Phi_t^{B}$. This property is useful in studying the non-Markovian properties of the channel and qualitatively comparing the two different measures we are interested in. The explicit form of the decoherence functions is not relevant for studying the behaviour qualitatively and is therefore presented in more detail in Sec.~\ref{sdc}.

\subsection{BCM measure}

Recently, a non-Markovianity measure based on monitoring the monotonicity of the quantum channel capacity $Q$ was introduced by Bylicka, Chruscinski and Maniscalco (BCM) \cite{channelcap}. The quantum capacity measures the ability of a quantum channel to reliably transmit information. For degradable channels, as the dephasing channels considered in this paper, $Q$ is defined in terms of the coherent information as follows
\begin{align}
Q\{\Phi_t\} = \underset{\rho}{\text{sup}} \hspace{2pt} I_c (\rho, \Phi_t).
\end{align}
Then based on $Q$, the measure is defined as
\begin{align}\label{BCM}
	\mathcal{N}_{BCM} = \int_{\frac{d Q\{\Phi_t\}}{d t}>0} Q\{ \Phi_t\} \text{d} t.
\end{align} 
The measure monitors and adds up the possible temporary increases in the capacity $Q$ to get a value such that any channel which has $\mathcal{N}_{BCM}>0$ is defined to be non-Markovian. In this work we will not calculate explicit values for the measure but only use the fact that as long as the capacity $Q$ is non-monotonic, the channel will be non-Markovian in the sense of the BCM measure.

\begin{figure}[t]
\includegraphics[width=0.47\textwidth]{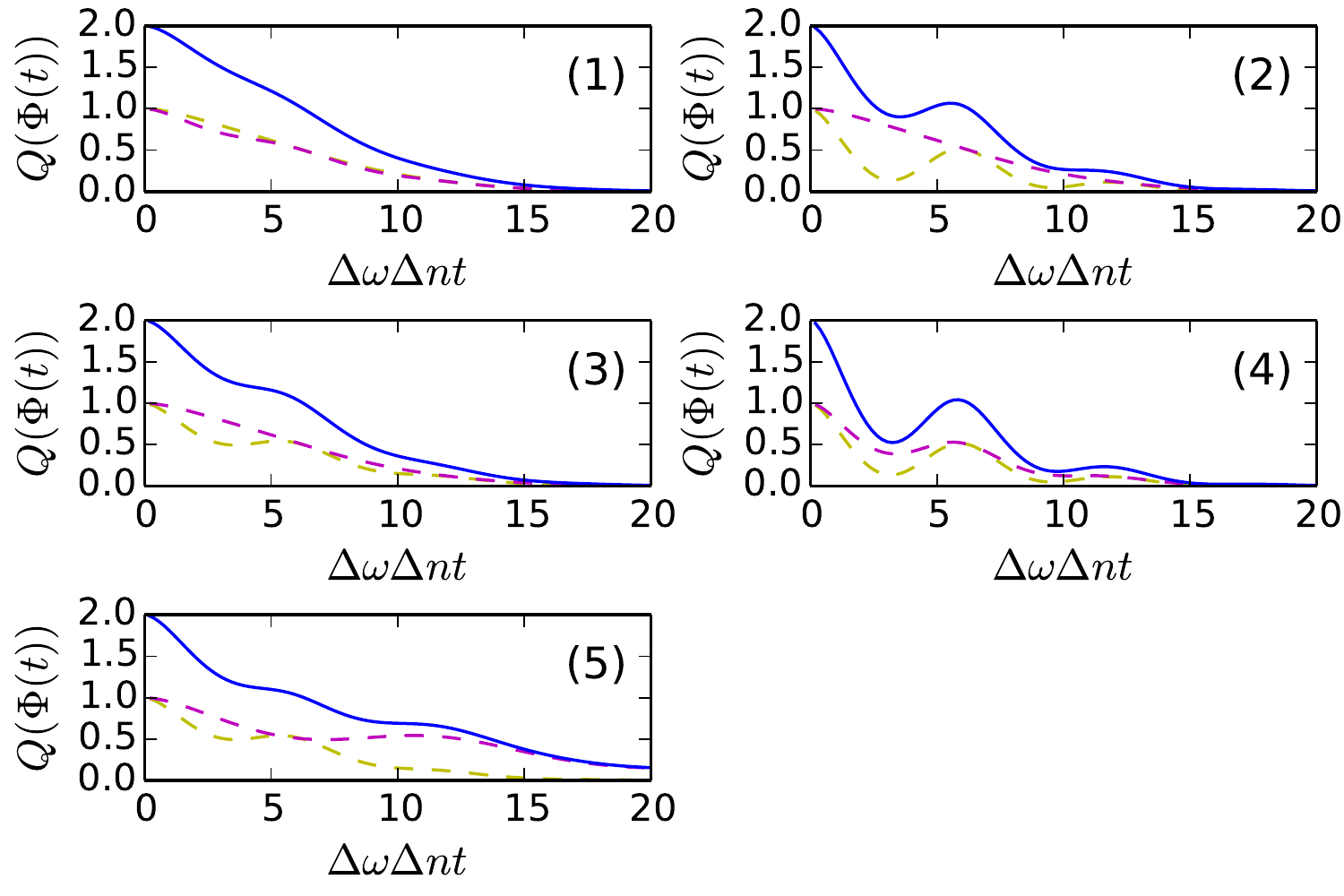}
\caption{The behavior of the quantum capacity for different combinations of different kinds of local channels. The dashed lines represent the local and the blue lines the global channels. The different combinations are tabulated in Table~\ref{locglob}.}
\label{qcaps}
\end{figure}

Using the method from \cite{chains}, $Q(\{ \Phi^{AB}_t \} )$ of the channel considered in our work can be written as
\begin{equation}\label{2noise}
Q( \{ \Phi^{AB}_t \}) =  2 - H_2\Bigg[\dfrac{1+ \vert \kappa_1(t)\vert}{2}\Bigg] - H_2\Bigg[\dfrac{1+ \vert \kappa_2(t)\vert}{2}\Bigg],
\end{equation}
where $H_2$ is the binary entropy function.  For uncorrelated local dephasing channels, the bipartite quantum channel capacity is additive: $Q(\{\Phi_t^{AB}\}) = Q(\{\Phi_t^{A}\}) + Q(\{\Phi_t^{B}\})$. Using this result and choosing as examples five different combinations of local dephasing channels, as listed in Table~\ref{locglob}, we  see different types of behavior for the global channel. The Table also contains the parameters $A^A$ and $A^B$ related to the experimental realisation of the Markovian and non-local channels in the photonics set-up (see Sec.~\ref{sdc}). The behavior of the capacities in the photonic realization is plotted in Fig.~\ref{qcaps}. The additivity property in \eqref{2noise} tells directly that combination of any two local Markovian dephasing channels always leads to a Markovian global channel. It also implies that using identical non-Markovian dephasing channels on both Alice's and Bob's side always leads to a global non-Markovian channel. However, we can also combine a Markovian and a non-Markovian channel to get both Markovian and a non-Markovian global channel. It is also possible to combine two different non-Markovian channels to get a Markovian global channel in sense of the BCM measure. 

%\begin{table}[h!]
\begin{table}[t]
\centering \label{combs}
\begin{tabular}{|c|c|c|c|c|c|}
\hline 
Combination & $A^A$ & $\Phi^A$ & $A^B$ & $\Phi^B$ & $\Phi^{AB}$ \\ 
\hline 
1 & 0.004 & M. & 0.026 & M. & M. \\ 
\hline 
2 & 0.377 & non-M. & 0.004 & M. & non-M. \\ 
\hline 
3 & 0.091 & non-M. & 0.004 & M. & M. \\ 
\hline 
4 & 0.377 & non-M. & 0.145 & non-M. & non-M. \\ 
\hline 
5 & 0.091 & non-M. & 0.091 & non-M.* & M. \\ 
\hline 
\end{tabular} 
\caption{Summary of different kinds of behaviors for the global channel with different parameter values. Here M. stands for Markovian in the sense of BCM measure. In the last case marked with * we have re-scaled the time parameter $t$ to $0.5t$ on Bob's side to get the oscillations of the decoherence functions out of phase. This amounts to changing the birefringence on Bob's side, as seen in the photonic implementation in Sec.~\ref{sdc}.}
 \label{locglob}
\end{table}

The combinations 3 and 5, plotted in Fig.~\ref{qcaps}, are particularly interesting. In the case of combination 5 two local non-Markovian 1-qubit channels give rise to Markovian 2-qubit channel with respect to the BCM measure. Because of this, the two independent channels can complement each other, which enables the above combination of two independent, locally non-Markovian channels to become globally Markovian. On the other hand, in the case of combination 3, one channel is Markovian enough to smooth out the non-Markovian behavior of the other channel  hence making the global channel Markovian. As we will see below, this cannot happen for the BLP measure.

\subsection{BLP measure}
Another way of defining Markovianity is to use trace distance, which is a metric defined by the trace norm on the set of quantum states. The trace distance of two quantum states $\rho_1$ and $\rho_2$ is defined as
\begin{align}
D(\rho_1, \rho_2) = \frac{1}{2}|| \rho_1- \rho_2 ||_{\text{tr}}.
\end{align}
$D$ is monotonic under PT maps \cite{ruskai} and also has a physical interpretation as it is closely related to the optimal probability $P_d(\rho_1,\rho_2)$ of distinguishing two unknown quantum states $\rho_1$ and $\rho_2$. The relation is
\begin{align}
P_d(\rho_1,\rho_2) = \frac{1}{2}(1+D(\rho_1,\rho_2)).
\end{align}
With this connection, an increase in the trace distance between pairs of states of a system of interest is interpreted as information flowing back into the system. The corresponding non-Markovianity measure, introduced by Breuer, Laine, and Piilo (BLP) is defined as \cite{NMprl} 
\begin{align}
\mathcal{N}_{BLP} = \underset{\rho_1(0),\rho_2(0)}{\text{sup}} \int_{\sigma(\rho_1(t), \rho_2(t))>0} \sigma(\rho_1(t), \rho_2(t) )\hspace{2pt} \text{d} t,
\end{align}
where 
\begin{align}
\sigma(\rho_1(t), \rho_2(t) )=\frac{d D(\rho_1(t), \rho_2(t))}{d t}.
\end{align}
The measure is built by adding up the increases in the trace distance between pairs of states during the evolution. Then this number is maximized over all possible choices for the initial states to get a quantity which characterises only the properties of the channel. Specifically, whenever $\mathcal{N}_{BLP}>0$ the channel is defined as non-Markovian.

%%%%%%%%%%%% FIG 1
%%%%%%%%%%%% FIG 1

The BLP measure of non-Markovianity can always detect the local non-Markovian behavior in the case of independent local channels. This can be done by choosing specific product states as the initial probe states and using the properties of the trace distance. Let $\rho_i^A$ be arbitrary states of Alice's system and $\rho_1^B$ an arbitrary state of Bob's system. Then
\begin{align}
D(\rho_1^A(t) \otimes \rho_1^B(t), \rho_2^A(t) \otimes \rho_1^B(t))=D(\rho_1^A(t), \rho_2^A(t)),
\end{align}
which shows that this choice of initial states is sensitive only to what happens on Alice's side. For example, let the initial states be
\begin{align}\label{states12}
\rho_1^{AB} &= \frac12 \begin{pmatrix}
1 & 1 \\ 
1 & 1
\end{pmatrix} \otimes \frac12 \begin{pmatrix}
1 & 0 \\ 
0 & 1
\end{pmatrix} =
\begin{pmatrix} 
\frac{1}{4} & 0 & \frac{1}{4} & 0 \\ 
0 & \frac{1}{4} & 0 & \frac{1}{4} \\
\frac{1}{4} & 0 & \frac{1}{4} & 0 \\
0  & \frac{1}{4} & 0 & \frac{1}{4} 
\end{pmatrix}, \\
\rho_2^{AB} &= \frac12 \begin{pmatrix}
1 & 0 \\ 
0 & 1
\end{pmatrix} \otimes \frac12 \begin{pmatrix}
1 & 0 \\ 
0 & 1
\end{pmatrix} =
\begin{pmatrix} 
\frac{1}{4} & 0 & 0 & 0 \\ 
0 & \frac{1}{4} & 0 & 0 \\
0 & 0 & \frac{1}{4} & 0 \\
0  & 0 & 0 & \frac{1}{4} 
\end{pmatrix}.
\end{align}
Now after mapping $\rho_1^{AB}$ and $\rho_2^{AB}$ with $\Phi^{AB}_t$ we get
\begin{align}
\Phi^{AB}_t(\rho_1^{AB}) - \Phi^{AB}_t(\rho_2^{AB}) &= 
\begin{pmatrix} 
0 & 0 & \frac{\kappa_1(t)}{4} & 0 \\ 
0 & 0 & 0 & \frac{\kappa_1(t)}{4} \\
\frac{\kappa^*_1(t)}{4} & 0 & 0 & 0 \\
0  & \frac{\kappa^*_1(t)}{4} & 0 & 0 
\end{pmatrix}
,
\end{align}
eigenvalues of which are $-\frac{\vert\kappa_1(t)\vert}{4}, -\frac{\vert\kappa_1(t)\vert}{4}, \frac{\vert\kappa_1(t)\vert}{4}$ and $\frac{\vert\kappa_1(t)\vert}{4}$. Thus the trace distance can be calculated as 
\begin{align}\label{d12}
D(\rho_1^{AB}(t), \rho_2^{AB}(t)) =&
 \dfrac{1}{2}\Bigg(\frac{\vert\kappa_1(t)\vert}{4}+\frac{\vert\kappa_1(t)\vert}{4}+\frac{\vert\kappa_1(t)\vert}{4}+\frac{\vert\kappa_1(t)\vert}{4}\Bigg)  \nonumber \\
=& \frac{\vert\kappa_1(t)\vert}{2}.
\end{align}

Since $\vert\kappa_1(t)\vert$ can be chosen to be non-monotonic (the case where the local channel on Alice's side is non-Markovian), so can the trace distance \eqref{d12}. Because the measure was defined as a maximization over all initial state pairs, this particular choice gives a lower bound for it. This implies that the dynamics of the system is non-Markovian with respect to BLP measure. Similar result can of course be obtained also for the case of a non-Markovian channel on Bob's side.  Thus it is clear that the BCM and BLP measures are not equivalent in the case of 2-qubit dephasing channels. Similar reasoning applies to any number of independent qubit channels. However, we see that different kinds of local behavior can lead to non-Markovian global behavior, meaning that the information transmission capacity of the two qubit channel is not necessarily only deteriorating over time. In the following section we will study the performance of our channel in the superdense coding protocol with different combinations of local noise.

\section{Two-qubit dephasing channel in the SDC protocol}\label{sdc}

Superdense coding is one of the best known examples of using entanglement for quantum information processing purposes \cite{bibsdc}. In the protocol Alice and Bob share one of the Bell states. Then Alice  applies a unitary transformation to her qubit to change the overall state to any of the four Bell states. Subsequently, she sends her qubit to Bob, who performs a measurement to find out the overall state. Because the states are orthogonal, they can be distinguished perfectly and thus four different messages can be sent from Alice to Bob with perfect fidelity. This equals a capacity of two classical bits with only one qubit and one bit of entanglement. 

\begin{figure}[t]
\includegraphics[width=0.47\textwidth]{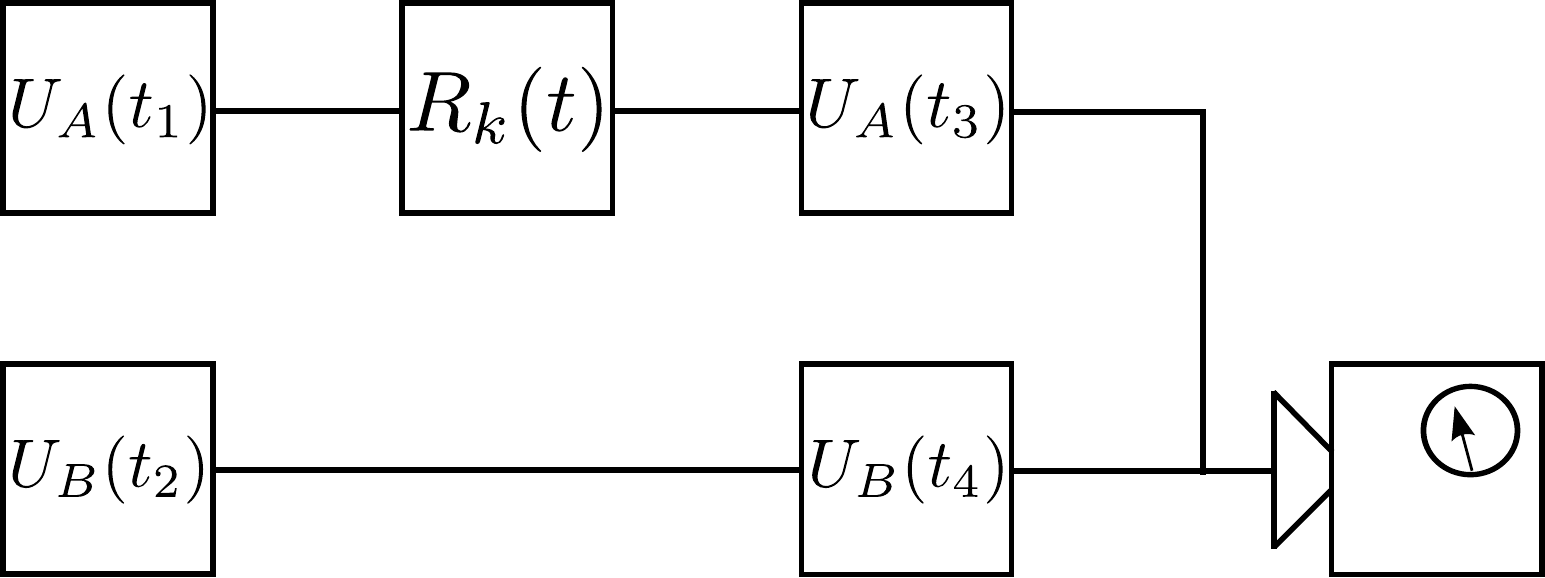}
\caption{Illustration of superdense coding scheme with local dephasing noises on Alice's and Bob's side. Each local noise $U_{A,B}(t)$ can be chosen independently of the other to examine different strategies to improve mutual information.
}
\label{beauty}
\end{figure}

Suppose Alice and Bob initially share two polarization entangled photons in the Bell state $\ket{\Phi^+}$, which they plan to use for the superdense coding protocol. However, in addition to the encoding operation on Alice's side, both photons are subjected to local, independent dephasing channels caused by unitary coupling between the polarization and frequency degrees of freedom. The frequency degree of freedom for Alice's and Bob's photons are characterized by the frequency distribution $g(\omega^A,\omega^B)$ which is normalized so that $\int d\omega^A d\omega^B \vert g(\omega^A,\omega^B)\vert^2 = 1$. The state of the combined system is thus
\begin{align}\label{init}
\ket{\Phi^+}\otimes\ket{\chi} =& \dfrac{1}{\sqrt{2}} (\ket{HH} + \ket{VV}) \\ \nonumber
&\otimes \int d\omega^A d\omega^B g(\omega^A,\omega^B)\ket{\omega^A \omega^B}.
%\otimes\ket{\omega^B}.
\end{align}
The couplings are  of the form \cite{nonlocal}
\begin{align}\label{2mutualuni}
&U_j(t) = \int d\omega^j (e^{i\omega^j n_V^j t} \ket{V}\bra{V} + e^{i\omega^j n_H^j t} \ket{H}\bra{H}) \otimes \ket{\omega^j}\bra{\omega^j}.
\end{align}
The channel structure is illustrated in Fig.~\ref{beauty}. First local dephasing noise $U_A(t_1)$ and $U_B(t_2)$ act both on Alice's and Bob's photons. After that Alice applies unitary encoding by using a local unitary $R_k(t)$ operation on her photon. The unitary matrix $R_k(t)$ is a modified Pauli matrix, used to make the decoherence function real. In order to achieve this, the interaction times must be known, so that $R_k(t)$ can be chosen accordingly. This amounts to tuning and calibrating the possible experimental realization of the superdense coding. We define the encoding operators $R_k(t)$ corresponding to the four possible messages as
\begin{align}\nonumber
R_0(t) &= \alpha^*(t)\beta^*(t)\ket{H}\bra{H} + \alpha(t)\beta(t)\ket{V}\bra{V}, \\ \nonumber
R_1(t) &= \beta^*(t)\ket{H}\bra{V} + \beta(t)\ket{V}\bra{H},\\ \nonumber
R_2(t) &= -i\beta^*(t)\ket{H}\bra{V} + i\beta(t)\ket{V}\bra{H},\\ \label{R}
R_3(t) &= \alpha^*(t)\beta^*(t)\ket{H}\bra{H} - \alpha(t)\beta(t)\ket{V}\bra{V}.
\end{align}
Here, $\alpha(t)$ and $\beta(t)$ are some time dependent complex functions such that $\vert\alpha(t)\vert = \vert\beta(t)\vert = 1 \hspace{0.2cm} \forall t$. A simple calculation shows that applying each $R_k(t)$ to the initial system state $\ket{\Phi^+}$ creates four orthogonal states which thus can be perfectly distinguished. The different messages and corresponding measurements are listed in Table~\ref{encoding}.
\begin{table}[t]
\centering
\begin{tabular}{|c|c|c|c|}
\hline
Input $x$                   & Alice performs                      & Bob measures                        & Output $y$\\ \hline
0                       & $R_0(t)$                      & $\ket{\Phi^+}$                      & 0   \\ \hline
1                       & $R_1(t)$                      & $\ket{\Psi^+}$                      & 1   \\ \hline
2                       & $R_2(t)$                      & $\ket{\Psi^-}$                      & 2   \\ \hline
3				   & $R_3(t)$				   & $\ket{\Phi^-}$ 				  & 3   \\ \hline
\end{tabular}
\caption{Alice's encoding and Bob's measurement operations for transmitting and receiving the messages 0-3.}
\label{encoding}
\end{table}
After Alice's encoding, local dephasing channels $U_A(t_3)$ and $U_B(t_4)$ act on both Alice's and Bob's qubit respectively. The operator describing the evolution of the total system for a given encoding operator $k$ takes the form
\begin{align}\label{2mutualtot}
U_k=U_A(t_3)R_k(t) U_A(t_1) \otimes U_B(t_4)U_B(t_2).
\end{align}
We assume that Alice and Bob can control the interaction times of their local noises.
For the following analytical calculations we assume that $t_1 = t_2 = t_3 = t_4 :=t/2$, i.e., all the four noises have the same duration and the total interaction times in Alice's and Bob's side are equal to $t$. Later on, we also present results when there is no noise on Bob's side and for the case when the only noise is that of Alice after her encoding.
When all four interactions are on, we can now calculate how the initial state \eqref{init} of the total closed system evolves when using different $R_k(t)$. Tracing out the environmental degrees of freedom gives the following
open system states with different encoding operations 
\begin{align}\label{states}\nonumber
\rho_0^S(t) =& \frac{1}{2}\big(\ket{HH}\bra{HH} + \ket{VV}\bra{VV} \\ \nonumber
&+\alpha(t)^2\beta(t)^2k^*(t)\ket{VV}\bra{HH} \\ \nonumber
&+ \alpha^*(t)^2\beta^*(t)^2k(t)\ket{HH}\bra{VV}\big),\\ 
\rho_1^S(t) =& \frac{1}{2}\big(\ket{VH}\bra{VH} + \ket{HV}\bra{HV} \\ \nonumber &+\beta(t)^2h^*(t)\ket{VH}\bra{HV} \\ \nonumber
&+ \beta^*(t)^2h(t)\ket{HV}\bra{VH}\big),\\ \nonumber
\rho_2^S(t) =& \frac{1}{2}\big(\ket{VH}\bra{VH} + \ket{HV}\bra{HV}  \\ \nonumber &-\beta(t)^2h^*(t)\ket{VH}\bra{HV} \\ \nonumber 
&-\beta^*(t)^2h(t)\ket{HV}\bra{VH}\big),\\ \nonumber
\rho_3^S(t) =& \frac{1}{2}\big(\ket{HH}\bra{HH} + \ket{VV}\bra{VV} \\ \nonumber &-\alpha(t)^2\beta(t)^2k^*(t)\ket{VV}\bra{HH} \\ \nonumber
&- \alpha^*(t)^2\beta^*(t)^2k(t)\ket{HH}\bra{VV}\big),
\end{align}
where 
\begin{align*}
h(t) &= \int d\omega^A d\omega^B e^{i\omega^B(n_V^B - n_H^B)t}\vert g(\omega^A,\omega^B) \vert^2, \\
k(t) &=\int d\omega^A d\omega^B e^{i\omega^A(n_H^A - n_V^A)t}e^{i\omega^B(n_H^B - n_V^B)t} \vert g(\omega^A,\omega^B) \vert^2
\end{align*}
and the subindex of $\rho_k$ specifies which $R_k$ was used to evolve the initial state (\ref{init}).

Alice and Bob can freely choose the form of $\vert g(\omega^A,\omega^B) \vert^2$ in the experimental realization. We are interested in the case of independent noise channels, which means that the joint frequency distribution $|g(\omega^A, \omega^B)|$ is a product distribution 
\begin{align}
\vert g(\omega^A,\omega^B) \vert = \vert g_A(\omega^A) \vert \vert g_B(\omega^B) \vert.
\end{align}
Suppose they agree on using a product of two double-peaked Gaussian distributions. The peaks of the Gaussians are centered at $\omega_1^j$ and $\omega_2^j$. Using this we can evaluate the integrals  as \cite{nonlocal}
\begin{align}\nonumber
h(t) &= \frac{e^{-\frac{1}{2}(\sigma\Delta n t)^2}}{1 + A^B}\Big(e^{i\Delta n \omega_1^B t} + A^Be^{i\Delta n \omega_2^B t}\Big), \\
k(t) &= \frac{e^{-\frac{1}{2}(\sigma\Delta n t)^2}}{1 + A^A}\Big(e^{i\Delta n \omega_1^A t} + A^Ae^{i\Delta n \omega_2^A t}\Big)h(t),
\end{align}
where $\Delta n = n_H^A - n_V^A = n_H^B - n_V^B$, $\sigma$ is the width of the peaks and $A^j =A_2^j/A_1^j$ is the relation  of the amplitudes of frequency peaks of photon $j$. By manipulating $A^j$ we can control whether the local environment is Markovian or non-Markovian \cite{NMNP}.

By choosing the complex functions used in unitary coding as 
\begin{align}\nonumber
\alpha(t)^2 &= \sqrt{\frac{e^{i\Delta n(2\omega_1^A + \omega_2^A)t} + e^{i\Delta n(2\omega_2^A + \omega_1^At)}A^A}{e^{i\Delta n\omega_2^At} + e^{i\Delta n\omega_1^At}A^At}}, \\
\beta(t)^2 &= \sqrt{\frac{e^{i\Delta n(2\omega_1^B + \omega_2^B)t} + e^{i\Delta n(2\omega_2^B + \omega_1^Bt)}A^B}{e^{i\Delta n\omega_2^Bt} + e^{i\Delta n\omega_1^Bt}A^B}}
\end{align}
\noindent we see that the decoherence functions $\beta^*(t)^2h(t) = \vert h(t) \vert$ and $\alpha^*(t)^2\beta^*(t)^2k(t)= \vert k(t) \vert$ become real valued. Using these choices, the final states that Bob obtains before his measurement become
\begin{align}\label{statesbm}\nonumber
\rho_0^S(t) =& \frac{1}{2}\big(\ket{HH}\bra{HH} + \ket{VV}\bra{VV} \\ \nonumber
&+|k(t)|\ket{VV}\bra{HH} \\ \nonumber
&+ |k(t)|\ket{HH}\bra{VV}\big),\\ 
\rho_1^S(t) =& \frac{1}{2}\big(\ket{VH}\bra{VH} + \ket{HV}\bra{HV} \\ \nonumber &+|h(t)|\ket{VH}\bra{HV} \\ \nonumber
&+ |h(t)|\ket{HV}\bra{VH}\big),\\ \nonumber
\rho_2^S(t) =& \frac{1}{2}\big(\ket{VH}\bra{VH} + \ket{HV}\bra{HV}  \\ \nonumber &-|h(t)|\ket{VH}\bra{HV} \\ \nonumber 
&-|h(t)|\ket{HV}\bra{VH}\big), \\ \nonumber
\rho_3^S(t) =& \frac{1}{2}\big(\ket{HH}\bra{HH} + \ket{VV}\bra{VV} \\ \nonumber &-|k(t)|\ket{VV}\bra{HH} \\ \nonumber
&- |k(t)|\ket{HH}\bra{VV}\big).
\end{align}
Before going further, we introduce tools for quantifying the performance of   the two-qubit channel in the superdense coding protocol.

\subsection{Mutual information}
Mutual information measures correlations between two random variables $X$ and $Y$. Basically it tells how much one can deduce from $Y$ by knowing $X$. In this sense, it is a natural measure to quantify the success of a messaging protocol where Alice wants to send a message to Bob. 

For two discrete random variables $X$ and $Y$ with the joint distribution $p(x,y)$ and marginal distributions $p(x)$ and $p(y)$, the classical mutual information is defined as 
\begin{align}
I(X:Y) = H\left(\{p(x)\}\right) + H\left(\{p(y)\}\right) - H\left(\{p(x,y)\}\right),
\end{align}
where $H$ is the Shannon entropy. By using the definition of Shannon entropy and the relation $p(x,y) = p(y\vert x)p(x) $ we see that
\begin{align}\label{probmutu}
I(X:Y) = \sum_{x\in X}p(x)\sum_{y\in Y}p(y\vert x)\log_2\frac{p(y\vert x)}{p(y)}.
\end{align}

Now let $X$ be the set of messages used by Alice and $Y$ the set used by Bob. Then $p(x)$ is the probability that Alice sends the message $x$ and $p(y)$ is the probability that Bob receives the message $y$.
$p(y\vert x)$ is the conditional probability of Bob receiving message $y$ given that Alice sent message $x$. In the superdense coding protocol the conditional probabilities can be calculated as
\begin{align}
p(y\vert x) = \text{tr}[E_y\rho_x],
\end{align}
where $\rho_x$ is the state that Alice encodes the message $x$ to and $E_y$ is the POVM element representing the measurement outcome associated to the message $y$ by Bob. For simplicity we assume a uniform distribution on Alice's messages, which means that $p(x)=p(y)=\frac{1}{4}$.

\subsection{Channel performance in terms of mutual information}

By using the reduced density matrices $\rho_k^S(t)$ defined in Eq. (\ref{statesbm}) we can obtain the conditional probabilities $p(y\vert x)$ of Eq. \eqref{probmutu} and then calculate the mutual information. For example, the conditional probability of Bob getting the incorrect result $\ket{\Psi^-}$ when Alice has performed the encoding $R_1(t)$ is
\begin{align}\nonumber
p(2\vert 1) &= \text{tr}\Big[\ket{\Psi^-}\bra{\Psi^-}\rho_1^S(t)\Big] \\ \nonumber
&= \frac{1-\vert h(t)\vert}{2}.
\end{align}
In a similar way one calculates also the other conditional probabilities. Combining these with the known probabilities $p(x) = p(y) = \frac{1}{4} \hspace{0.2cm}\forall x \in X, y \in Y$ we get for the mutual information
\begin{align}\nonumber
I&(X:Y) = \sum_{x=0}^3\frac{1}{4}\sum_{y=0}^3p(y\vert x)\log_2\frac{p(y\vert x)}{1/4} \\ 
& = 2 - \frac{1}{2}\Bigg{\{} H_2\Bigg[\frac{1+\vert k(t) \vert}{2}\Bigg] + H_2\Bigg[\frac{1+\vert h(t) \vert}{2}\Bigg] \Bigg{\}}.
\label{mutu2qubit}
\end{align}

Interestingly, the time-dependent mutual information in \eqref{mutu2qubit} is almost the same as the quantum channel capacity calculated in \eqref{2noise}. In the following, we examine different possibilities for dynamics of mutual information by plugging different noise configurations of local dephasing channels into \eqref{mutu2qubit}. It is easy to see that if $\vert k(t) \vert$ and $\vert h(t) \vert$ are monotonic, so is the mutual information. On the other hand if $\vert k(t) \vert$ and $\vert h(t) \vert $ both have recoveries at the same time intervals $t_1\le t \le t_2$, then also the mutual information has recoveries at the same intervals.

\begin{figure}[t]
\includegraphics[width=0.47\textwidth]{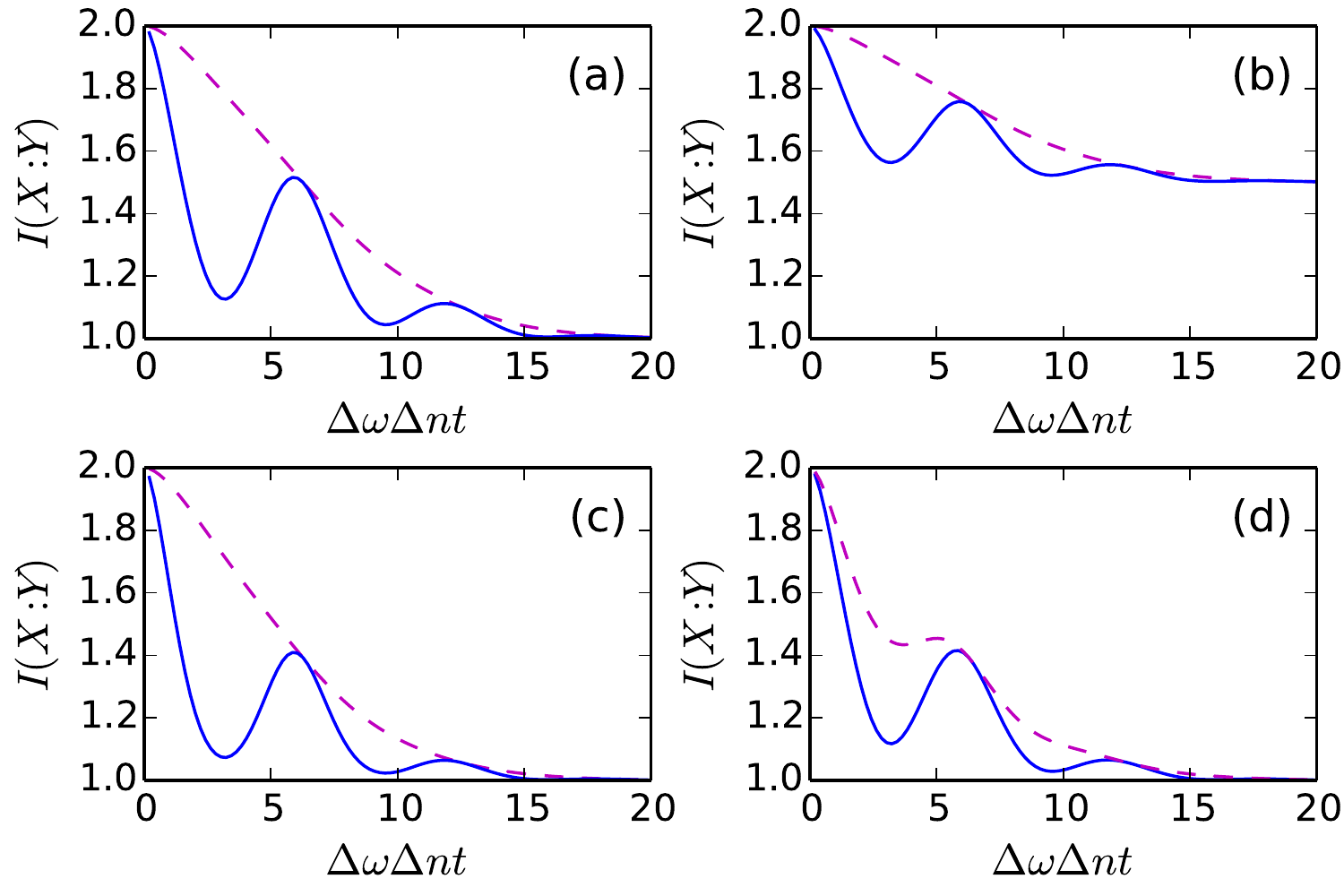}
\caption{
Mutual information as a function time in four different noise configurations.
(a) Noise only on Alice's side after her encoding.
(b) Noise only on Alice's side before and after her encoding. 
(c) Noise on Alice's side before and after her coding and identical noise on Bob's side. 
Here, in (a)-(c), solid blue line corresponds to non-Markovian local and dashed purple to Markovian local channels.
For panel (d), there is noise on Alice's side before and after her encoding and different noise on Bob's side. Here, solid blue line corresponds to Markovian local channel on Alice's side and non-Markovian local channel on Bob's side while the dashed purple line corresponds to the opposite case.}
\label{qmutus}
\end{figure}

Figure~\ref{qmutus} shows the behavior of mutual information for four different
noise configurations. We use fixed parameter values of $\sigma^A = \sigma^B = 1.8\times10^{12}$ Hz, $\Delta\omega^A = \Delta\omega^B  = 1.6\times10^{16}$ Hz and $\Delta n^A = \Delta n^B = \Delta n$. Two different local channels are used, Markovian and non-Markovian with respect to both BLP and BCM measure. The Markovian one corresponds to the choice of parameter $A^j = 0.004$ and the non-Markovian one to  $A^j = 0.390$. 

%%%%%%%%%% FIG 3
%%%%%%%%%% FIG 3

Figure~\ref{qmutus} (a) compares the dynamics of mutual information between Markovian and non-Markovian cases when there is noise only in Alice's side after her encoding. As expected, memory-effects revive the mutual information temporarily and finally the value approaches the classical limit equal to 1 in both cases. However, when we add noise also before Alice's encoding on her side, this improves the situation both for Markovian and non-Markovian cases, see Fig.~\ref{qmutus} (b). In addition of the slower decrease of mutual information and revivals, it is very interesting to notice that the asymptotic values approach now 1.5 which is significantly higher than the classical limit 1. When there is no noise in Bob's side and the duration of the noise is equal before and after Alice's encoding, then $|h(t)|=1$, and Eq.~\eqref{statesbm} shows that two states $\rho_1^S$ and $\rho_2^S$ fully recover their quantum features by an echo mechanism.  States $\rho_0^S$ and $\rho_3^S$, which depend on $|k(t)|$, eventually fully dephase. This leaves us three distinguishable cases with four encoding operations, and subsequently the value of mutual information remains higher than the classical limit, and at the same time below the value $\log_23$ if only three encoding operations were used.  

Figure~\ref{qmutus} (c) shows the results for the case having the same Markovian or non-Markovian noise on both sides of Alice and Bob. Here, the behaviour is very similar to
Fig.~\ref{qmutus} (a). The difference is quite obvious with stronger reduction and smaller revival of mutual information since added identical noise to Bob's side. The situation is more interesting when the noise applied in the two sides is different. Figure~\ref{qmutus} (d) shows the results when Alice has Markovian and Bob non-Markovian noise or viceversa.
Here, the values of mutual information are higher when the non-Markovian noise acts on Alice's side instead of Bob's side.  We conclude that the combination of Alice's encoding  operation with subsequent echo mechanism for two of the states and the non-Markovian character of her local channel is more efficient for SDC coding than placing the non-Markovian channel to Bob's side.
 
\section{Conclusions}

In this paper we have studied the capability of two non-Markovianity measures in quantifying memory effects for two independent dephasing channels.
The results for the BCM measure show that having at least one local non-Markovian channel can lead to both Markovian and non-Markovian global channel.
In contrast, the BLP measure always detects the local non-Markovian behavior of the global map in the case of independent channels.
It thus turns out that the BCM measure better captures the usefulness of the channel structure in transmitting information in the SDC protocol for the considered cases. We have further studied various dephasing noise configurations to optimize the information transmission. The results show that when noise affects only Alice's side, it is beneficial if it is present both before and after her encoding. In this case, the asymptotic limit of mutual information is significantly higher than the classical limit. 
Moreover, when noise is present in both Alice's and Bob's side -- one of them being Markovian and other one non-Markovian -- it is more useful for information transmission in SDC protocol to have non-Markovian channel on Alice's side.
Our results help in understanding how reservoir engineering and memory effects can be used to improve various quantum information based protocols.

\acknowledgments
This work has been supported by  the Magnus Ehrnrooth Foundation,
the EU Collaborative project QuProCS (Grant Agreement 641277), and the Academy of Finland (Project no. 287750).

\end{document}